\documentclass{iaus}
\usepackage{graphicx}

\title[JD 11.~~Pre-solar grains and AGB stars] 
{$\eta$~Carinae long-term variability}

\author[A.~Damineli, M.~Teodoro, M.~Corcoran \and J.~H.~Groh]   
{A. Damineli$^1$, M.~Teodoro$^1$, M.~Corcoran$^2$ \and J.~H.~Groh$^3$}

\affiliation{$^1$Instituto de Astronomia, Geof\'{\i}sica e Ci\^encias Atmosf\'ericas, Rua do Mat\~ao 1226, S\~ao Paulo, 05508-900, Brazil \\ email: {\tt damineli@astro.iag.usp.br} \\[\affilskip]
$^2$NASA/GSFC, Code 662, X-ray Astrophysics, Greenbelt, MD 20771, USA \\[\affilskip]
$^3$Max-Planck-Institute f\"ur Radioastronomie, Auf dem H\"ugel 69, D-53121 Bonn, Germany}

\pubyear{2010}
\volume{272}  
\pagerange{119--126}
\jname{Active OB stars}
\editors{C. Neiner, G. Wade, G. Meynet \& G. Peters}
\begin{document}

\maketitle

\begin{abstract}
We present preliminary results of our analysis on the long-term variations observed in the optical spectrum of the LBV star $\eta$~Carinae. Based on the hydrogen line profiles, we conclude that the physical parameters of the primary star did not change in the last 15 years.

\end{abstract}

\firstsection 
\section{Introduction}
Photometric monitoring of $\eta$~Carinae (\cite{feinstein67,feinsteinetal74,sterkenetal96,sterkenetal99,vangenderenetal06,frew04,fernandezlajusetal09}) revealed that an increase in brightness at variable rates, since 1950. The mechanism behind such long-term variations are still unclear, however. In this regard, spectroscopic monitoring can put several important constraints to the diagnosis. Unfortunately, frequent spectroscopic observations of this object began just about 2 decades ago, which is not sufficient yet to draw a clear picture of what is happening to the central source.

\section{Results and discussion}
In order to verify whether or not the central source in $\eta$~Car is passing through changes, we analized ground-based spectroscopic data taken at the same phase ($\phi\approx0.3$) of the spectroscopic event, but in different cycles (\#9, \#10, \#11 and \#12).

\begin{figure}[t]
\begin{center}
 \includegraphics[width=0.8\textwidth]{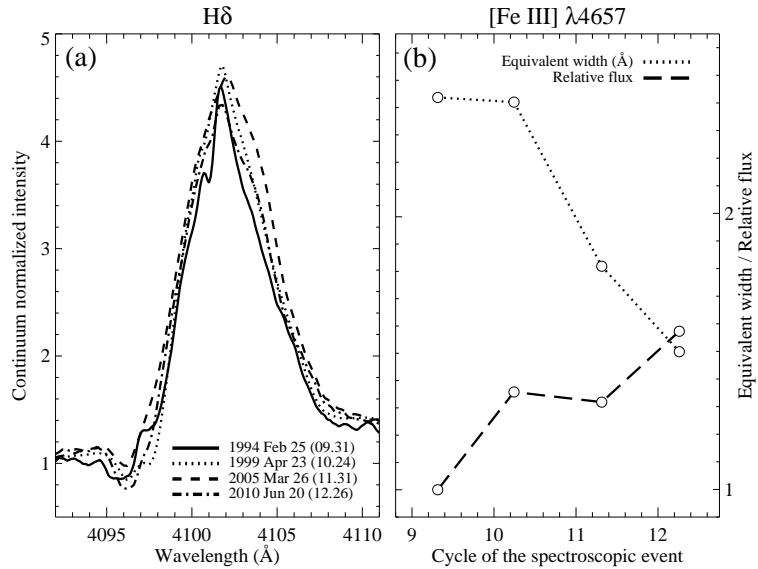} 
 \caption{(a) Line profile of H$\delta$ at the epochs indicated in the legend. (b) Equivalent width and relative line flux. The hydrogen line shows no significant variations throughout the last 15 years. On the other hand, the equivalent width of the [Fe\,{\sc iii}] emision line is systematically decreasing with time, but the relative line flux is increasing.}
   \label{fig1}
\end{center}
\end{figure}

Our analysis revealed that the lines formed in the wind of the primary star -- such as the hydrogen lines -- do not present any evidence of systematic or significant changes in line profile, as shown in Fig.\,\ref{fig1}a, for example. In that figure, the H$\delta$ line profile was normalized by the local continuum.

On the other hand, lines with high-ionization potential -- such as [Fe\,{\sc iii}]~$\lambda4657$ -- do show systematic variations, namely, the intensity of the peak of the line's narrow component is decreasing with time, relative to the local continuum and, thus, the equivalent width of the narrow component is \textit{decreasing} with time, as indicated in Fig.\,\ref{fig1}b (dotted line).

However, since forbidden lines are formed in a more extended region, and we do not know where exactly the increase in brightness is coming from, we converted the equivalent width measurements into line flux by using the $B$-band magnitudes for each epoch. After that, we normalized the line fluxes by the line flux observed in 1994 Feb 25th ($\phi=9.31$). The result is shown in Fig.\,\ref{fig1}b (dashed line).

After correcting the equivalent width of the narrow component of the [Fe\,{\sc iii}]~$\lambda4657$ emission line by the flux in the local continuum, the trend changed completely: the line flux is \textit{increasing} with time. From 1994 to 2010, the narrow component line flux increased by about 60 per cent (in the same period, the continuum flux increased by a factor of 2.5).

We know that the narrow component is formed in the Weigelt's blobs. If the equivalent width of such component is decreasing while the line flux is increasing, then we can conclude that either (1) the total extinction around the central source is decreasing in all directions, not only in our line-of-sight or (2) the effective temperature of the secondary star is increasing (or the wind opacity is decreasing).

Unfortunately, based only on our preliminary results shown in this proceedings, we can only conclude that the wind of the primary star did not change during the last 15 years. At this moment, we cannot point for sure which of the possibilities presented above is the correct one to explain the behavior of the [Fe\,{\sc iii}]~$\lambda4657$ line (although we favor the decreasing of extinction in all directions). However, further analysis of other spectral features with high-ionization potential will eventually provide us with more indications on what is happening in the central source of $\eta$~Car. That will be the subject of a more complete, forthcoming paper.

\end{document}